\documentstyle[prl,aps,multicol,epsfig,psfig,array,graphics]{revtex}

\newcommand{\sch}{Schr\"{o}dinger }
\newcommand{\eq}[1]{Eq.~(\ref{#1})}

\begin{document}

\title{Excited stationary states of trapped Bose-Einstein condensates}

\author{John A. Vaccaro, Ole Steuernagel and Ben Lorimer}
\address{Department of Physical Sciences, University of Hertfordshire,
Hatfield, Herts, AL10 9AB, UK
\\}
\date{\today}

\maketitle
\begin{abstract}
{We investigate the excited stationary states of Bose-Einstein condensates
trapped in harmonic potentials.  We derive simple analytical approximations of
the first few eigenstates of the associated time-independent one-dimensional
Gross-Pitaevskii equation and their energies. Our results are excited state
generalizations of the Thomas-Fermi approximation of the ground state. }
\end{abstract}

\begin{multicols}{2}

As opposed to traditional studies of liquid helium, the recent
experimental work on Bose-Einstein condensation \cite{expts}
centers on inhomogeneous mesoscopic systems. However, such
systems are highly nonlinear and no exact analytical expressions
for the wavefunctions and associate energies are yet known. In
order to understand the behavior of Bose-Einstein condensates
(BEC) it would be beneficial to have at least simple approximate
expressions for the condensate wave functions. These could be
used to study the qualitative behavior of the condensate and,
given the approximate expressions are precise enough, might even
be useful for the determination of overlap factors between
different wave functions, spectra and other quantities of
interest. Yukalov {\em et al.} \cite{yyb} have investigated this
problem using a perturbation method. Also, Kivshar {\em et al.}
\cite{kivshar} recently highlighted the connection between this
problem and that of dark solitons and gave numerical solutions to
the first few excited stationary states. In this communication we
derive simple analytical approximations to the stationary states
$\Phi_{n}$ $(n=0,1,2,\ldots )$ of the one-dimensional
Gross-Pitaevskii equation for a BEC confined in a harmonic trap
\cite{Perez98}:
\begin{equation}
\left( - {\frac {\partial ^{2}}{\partial z ^{ 2}}}\, + \lambda^2 \, z^{2}\, +
\frac{Q}{2} \,|\Phi_n (z)|^{2} - 2 \varepsilon_n \right) \Phi_n (z )=0 \; .
\label{bec_1}
\end{equation}
Here $z$ is the longitudinal trap coordinate, the aspect ratio, $\lambda=
\omega_z/ \omega\ll 1$, is given by the ratio of the longitudinal trap
frequency~$\omega_z$ to the transversal trap frequency~$\omega$, $\lambda$
formally is the effective spring constant of the 1-dimensional problem.
$\lambda$ is assumed to be much smaller than unity in order to yield a
cigar-shaped condensate which can be modeled by a 1-dimensional problem.
$Q/2=4\pi a N / a_0$ is the effective non-linearity for the interaction of the
$N$ trapped atoms with mass $m$ and scattering length $a$ in the trap with an
effective diameter $a_0=\sqrt{\hbar/(m \omega)}$~\cite{Perez98}.

The states $\Phi_n$ are normalized to unity, but, owing to the non-linearity of
the Gross-Pitaevskii-equation, they do not form an orthogonal set, generally
$\langle \Phi_{n}|\Phi_{m} \rangle \neq 0$.

It is convenient to rescale the Gross-Pitaevskii-equation to make the
coefficients of the kinetic and the non-linear term equal to unity.  Using the
rescaling $Y_n=\Phi_n \frac{1}{2} \sqrt{Q/ \varepsilon_0}$ and $x=z\sqrt{2
\varepsilon_0}$ Eq.\ (\ref{bec_1}) becomes
%
\begin{equation}
\left( - \,{\displaystyle {\frac {\partial ^{2}}{\partial x ^{ 2}}}}\, +
{\displaystyle  \lambda^2 \, x^{2}\,  + |Y_n ( x )|^{2} - \epsilon_n } \right)
\, Y_n ( x )  = 0  \; .
\label{bec_gpe}
\end{equation}
%
The advantage of this scaling is that the solutions are
parameterized solely by $\lambda$. The energy eigenvalue
$\epsilon_n = \varepsilon_n / \varepsilon_0$ is now scaled in
units of the Thomas-Fermi energy $\varepsilon_0=({3}^{2/3}/8)
\cdot (Q \lambda )^{2/3}$ of the lowest stationary state and the
spring constant is expressed in terms of $\lambda^2=
(\omega_z/\omega)^2 = (512/9)
\cdot \varepsilon_0^3/Q^2$.
Our choice of rescaling results in the normalization $\langle
\Phi_n |\Phi_n
\rangle=1= \frac{4\sqrt{\varepsilon_0}}{\sqrt{2}Q} \langle Y_n|Y_n \rangle $
which can be written as
%
\begin{equation}
\langle Y_n|Y_n \rangle = \frac{4}{3\lambda} \; .
\label{bec_norm}
\end{equation}
%

\begin{figure}
\psfig{figure={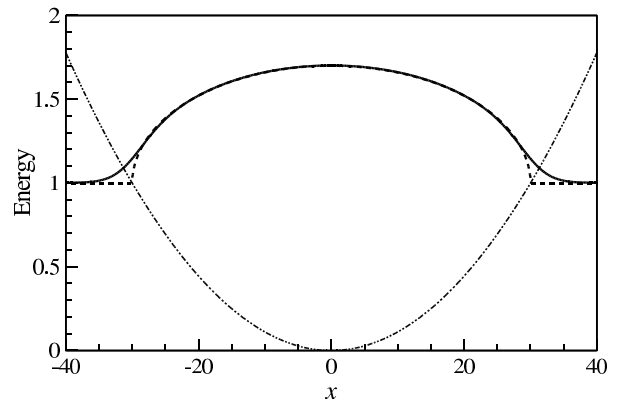},width=60mm,angle=-90,bbllx=0mm,bblly=10mm,bburx=205mm,
bbury=280mm,clip=}
\caption{Plot of the parabolic trap potential, the standard Thomas-Fermi
solution~{\protect \cite{baym}} for the condensate state $Y_0$ (dashed line)
and the corresponding exact numerical solution (solid line).  The value of the
nonlinearity is $\lambda = 1/30$. Both solutions have been displaced vertically
by their respective energy values and share the same arbitrary vertical
scaling.
} \label{fig1}
\end{figure}

The Thomas-Fermi approximation of $Y_0$ entirely neglects the
contribution from the kinetic energy term and yields the standard
approximate solution~\cite{baym} of (\ref{bec_gpe}):
%
\begin{equation}
Y_{0} (x) = \sqrt{1 - \lambda^2 \, x^2}, \mbox{   for } |x| < X_0 = 1/\lambda
\label{bec_tf0}
\end{equation}
%
and $Y_{0}(x)=0 $ otherwise, as illustrated in Fig. \ref{fig1}.
To generalize the Thomas-Fermi approach to excited stationary states, $Y_n$ for
$n>0$, we need to include the kinetic energy term.  Our approach to deriving a
solution is based on a piecewise ansatz which is motivated as follows. Over
small regions about some fixed point $x=x'$ the potential energy term
$\lambda^2 x^2$ can be considered slowly varying compared to the other terms in
Eq.~(\ref{bec_gpe}).  The local solution about $x'$ is then given approximately
by the solution of the nonlinear \sch equation
%
\begin{eqnarray}
\left( - \,{\displaystyle {\frac {\partial ^{2}}{\partial x ^{ 2}}}} + |Y_n ( x
)|^{2} - \epsilon'_n \right) \, Y_n ( x )  = 0  \; .
\label{bec_gpefix}
\end{eqnarray}
%
where $\epsilon'_n=\epsilon_n-\lambda^2 \, x'^{2}$. This equation is well known
in soliton theory;  the solutions with finite amplitude are given by the
Jacobi-sine function `sn':
%
\begin{equation}
\sqrt{ \epsilon'_n} \, A \, \mbox{ sn}\left(\sqrt{ \epsilon'_n (1-A^2/2)}
(x-a) , \frac{A}{\sqrt{2-A^2}}\right)
\label{sn}
\end{equation}
%
where $a$ is an arbitrary displacement along the $x$ axis and $0<A\leq 1$. The
wavelength of the function in (\ref{sn}) is given by
%
\begin{equation}
\Lambda=\frac{4}{\sqrt{\epsilon'_n (1-A^2/2)}} \times K(\frac{A}{\sqrt{2-A^2}})
\ , \label{lambda}
\end{equation}
%
where $K$ is a complete elliptic integral of the first kind \cite{Abramowitz}.
Imagine joining the solutions for two different values of $x'$ smoothly at some
intermediate point along the $x$ axis.  As an example, let this intermediate
point be a zero of the two solutions.  The slope of (\ref{sn}) at a zero is
given by $\epsilon'_n A\sqrt{1-A^2/2}$ and so a smooth join requires the value
of $A$ to be larger for the solution with the larger value of $|x'|$ to
compensate for the smaller value of $\epsilon'_n=\epsilon_n-\lambda^2 x'^2$.
Repeating this smooth joining of solutions for successive pairs of $x'$ values
along the $x$ axis leads to a set of piecewise solutions for which the value of
$A$ increases with $|x'|$.  At the critical value $A=1$ the wavelength becomes
infinite and the solution of \eq{bec_gpefix} is given by the corresponding
limit of~(\ref{sn}) namely
%
\begin{equation}
\sqrt{ \epsilon'_n}\, \tanh\left(\sqrt{\epsilon'_n/2}\,(x-a)\right) \ .
\label{tanh}
\end{equation}
%
This piecewise analysis implies that the general finite-amplitude solution of
\eq{bec_gpe} will be oscillatory near the origin and change to a tanh-like
function some distance away. Moreover, the $\tanh$ function in (\ref{tanh}) is
relatively flat for $|x-a|$ significantly different from zero, and so
(\ref{tanh}) is approximated well by just the prefactor $\sqrt{\epsilon'_n}=
\sqrt{\epsilon_n - \lambda^2 x'^2}$. We note that this prefactor is the
Thomas-Fermi solution~(\ref{bec_tf0}) in terms of the local variable $x'$.

 We can reduce the number of piecewise segments needed
in our ansatz using energy considerations~\cite{longpaper}. The potential
energy and its variation are smallest in the center of the trap and to a good
approximation we can neglect the trap potential in this region. This amounts to
replacing $\epsilon'_n$ with $\epsilon_n$ and holding $A$ constant in our
oscillatory solution (\ref{sn}) to give an oscillating function $Y_{\rm osc}$
which has a fixed wavelength and amplitude. Further away from the center of the
trap the potential energy increases at the expense of the kinetic energy to the
extent that the later becomes negligible. The wave function is then well
described by the conventional Thomas-Fermi solution~\cite{longpaper},
 $\sqrt{\epsilon_n -\lambda^2 x^2}$. Thus, replacing $x'$ with $x$ in
the prefactor $\sqrt{\epsilon_n - \lambda^2 x'^2}$ of (\ref{tanh}) yields a
function $Y_{\rm tf}$ which provides a transition between $Y_{\rm osc}$ and the
Thomas-Fermi solution.  We will call $Y_{\rm tf}$ the ``tail function'' as it
gives the shape of the condensate in its outer regions.

Our ansatz is to match these two segments $Y_{\rm osc}$ and $Y_{\rm tf}$
smoothly at points $x=\pm T_n$ which lie symmetrically about the origin. In
analogy to Eq.~(\ref{bec_tf0}) we also set $Y_n(x)=0$ for $|x|>
X=\sqrt{\epsilon_n} /\lambda$ where $X$ is the half-width of the condensate in
the Thomas-Fermi approximation. Thus our ansatz is given by
%
\begin{equation}
  Y_n(x) = \left\{ \begin{array}{ll}
                 Y_{\rm osc}(x), &  |x|\le T_n
                  \\
                 Y_{\rm tf}(x), &  T_n<|x|\le X
                  \\
                  0,            &  X < |x|
              \end{array} \right.
\label{bec_tf3}
\end{equation}
%
where
%
\begin{eqnarray*}
  Y_{\rm osc}(x) &=& \sqrt{ \epsilon_n}\, A \\
               & &\times \mbox{sn}\left( \sqrt{ \epsilon_n (1-A^2/2)}
                  (x-T_n) , \frac{A}{\sqrt{2-A^2}} \right)\ ,\\
  Y_{\rm tf}(x)&=& \left(\frac{x}{|x|} \right)^n \sqrt{ \epsilon_n - \lambda^2 \,x^{2}  }\\
               & &\times \tanh\left( \sqrt{(\epsilon_n -\lambda^2 T^2 )
                   /2 }\,(|x|-T_n)\right).
\end{eqnarray*}
%
The points $x=\pm T_n$, where the solution switches between the
oscillatory part and the tail function is given by $T_n=(n-1)\;
\Lambda/4$. The forms of $T_n$ and $Y_{\rm osc}(x)$ ensure that
$Y_n(x)$ is either symmetric or antisymmetric for $n$ even or
odd.

The parameters $A$, $T_n$ and $\epsilon_n$ are determined by the
requirements that the two segments match smoothly and that the
solution is normalized according to \eq{bec_norm}.  The first
requirement is satisfied when the gradients, ${\frac {\partial
}{\partial x}}\,Y_{\rm osc}|_{x=T_n}$ and $ {\frac {\partial
}{\partial x}}\,Y_{\rm tf}|_{x=T_n}$ are equal, i.e. when
%
\begin{equation}
  \epsilon_n A \sqrt{ 1 - A^{2}/2} =
  (\epsilon_n -\lambda^2\,T_n^{2})/\sqrt{2}\ .
  \label{gradients}
\end{equation}
%
For notational simplicity we concentrate on the second excited state $Y_2$,
i.e. $n=2$. The expression for $T_2$ can be approximated by $T_2\approx
\sqrt{2/\epsilon_2} $ arctanh$ (\sqrt{(1+A^2)/2})$ $ \approx
\sqrt{2/\epsilon_2} \cdot \mbox{ln}( 2/\sqrt{1-A}) $. This can be used to solve
Eq.~(\ref{gradients}) for $A \approx 1 - \lambda \;\epsilon_2\; {W}(8
\epsilon_2 / \lambda ) /2$ where $W$ is the Lambert function defined by $W
\exp(W) = x$. This, in turn, can be approximated to give $A \approx 1 - \lambda
\; \epsilon_2 \ln( 8 \epsilon_2/\lambda)/2$ and therefore $T_2 \approx - \ln (
\lambda\; \epsilon_2 \ln( 8 \epsilon_2/\lambda)/ 8) / \sqrt{2\epsilon_2}$. Thus
we now have a simple expression for $T_2$ in terms of $\epsilon_2$.

The second requirement, i.e. that $Y_2(x)$ be normalized according
to \eq{bec_norm}, determines the value of~$
\epsilon_2$. The contribution from the oscillatory part of the
wave function is
%
\begin{eqnarray*}
 \langle Y_{\rm osc} | Y_{\rm osc} \rangle &=& 2 \left[ \sqrt{2} \mbox{ arctanh}
 (\sqrt{(1+A^2)/2})\right.  \\
 & &\left. - \sqrt{(1+A^2)/\epsilon_2}\right]
\end{eqnarray*}
%
and the contribution from the tails is approximately
%
\begin{eqnarray}
 \langle Y_{\rm tf} | Y_{\rm tf} \rangle &\approx&
        2 \int_T^{X} dx \; \; \left[\lambda^2 (T^2-x^2) \right.
  \nonumber \\
  & &\left. + (\epsilon_n-\lambda^2 T^2) \tanh^2 (\sqrt{(\epsilon_n-\lambda^2 x^2)/2})\right].
\label{bec_n2tanh}
\end{eqnarray}
%
We use tanh($ \frac { \sqrt{ \epsilon } X }{\sqrt{2} }) \approx 1$ to find a
simplified expression for~(\ref{bec_n2tanh}). Inserting the above expression
for $A$ and $T$ into $\langle Y_2 | Y_2 \rangle \doteq 4 /(3 \lambda)$,
treating $\lambda$ and $\delta_2=\epsilon_2-\epsilon_0$ as small expansion
parameters and performing a somewhat tedious calculation using an expansion in
$\delta_2$ followed by an expansion in $\lambda$ finally yields
$\epsilon_2\approx 1+2\sqrt{2} \lambda$.

The derivation for the general case $n = 1,2,3,...$ is very similar. We find
that
%
\begin{eqnarray}
   \epsilon_n & \approx &  1 + n \sqrt{2} \; \lambda \; ,
   \label{bec_e3} \\
   A & \approx & 1 - \lambda \;\epsilon_n
   \ln( 8 \epsilon_n/\lambda)/2 \; ,
   \label{A} \\
 \mbox{and } \; \;  T_n & \approx & (1-n) \ln (\lambda\;
   \epsilon_n \ln( 8 \epsilon_n/\lambda)/ 8) / \sqrt{2
   \epsilon_n} \; .
   \label{T_n}
\end{eqnarray}
%
\eq{bec_tf3} together with Eqs.~(\ref{bec_e3}-\ref{T_n}) are the main results
of this communication and give our analytical approximations of the stationary
states of a harmonically trapped condensate. We note, however, that the greater
the value of $n$, the more stringent the requirements on the aspect ratio
$\lambda$ to be small.

\begin{figure}
\psfig{figure={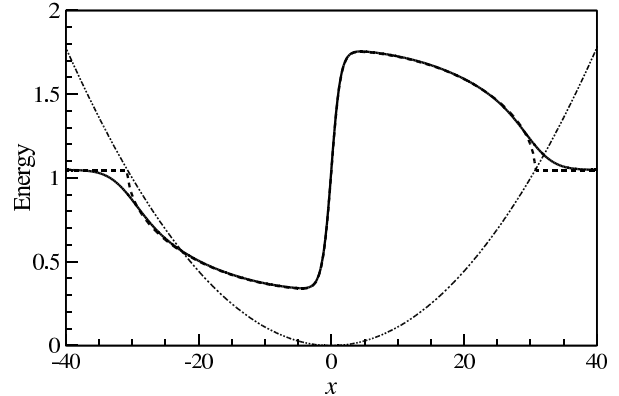},width=60mm,angle=-90,bbllx=0mm,bblly=10mm,bburx=205mm,
bbury=280mm,clip=}
\caption{Plot of the parabolic trap potential, our approximate analytical
solution for the condensate state $Y_1(x)$ (dashed line) and the corresponding
numerical solution (solid line) for a nonlinearity parameter of $\lambda =
1/30$.  As in Fig.~\ref{fig1} both solutions have been displaced vertically by
their respective energy values and they share the same arbitrary vertical
scaling.
} \label{fig2}
\end{figure}

\begin{figure}
\psfig{figure={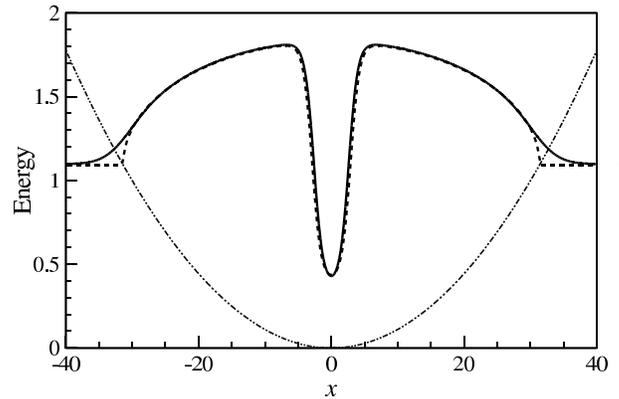},width=60mm,angle=-90,bbllx=0mm,bblly=10mm,bburx=205mm,
bbury=280mm,clip=}
\caption{Analogous to Fig.~\ref{fig2} for the second excited stationary state
$Y_2(x)$.
} \label{fig3}
\end{figure}

\begin{figure}
\psfig{figure={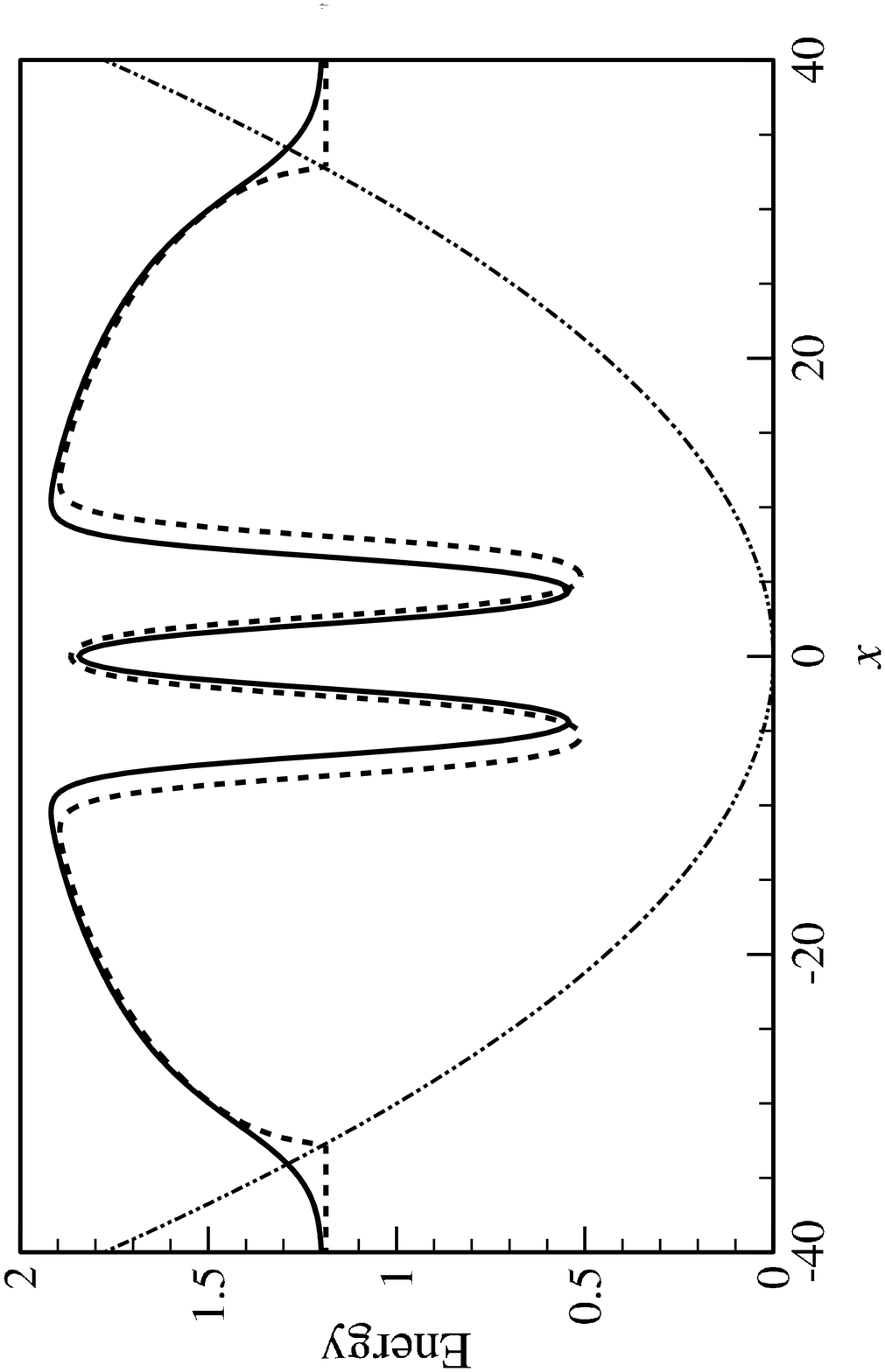},width=60mm,angle=-90,bbllx=0mm,bblly=10mm,bburx=205mm,
bbury=280mm,clip=}
\caption{Analogous to Fig.~\ref{fig2} for the fourth excited stationary state
$Y_4(x)$.
} \label{fig4}
\end{figure}

We should expect, and do obtain, best results for low order $n$ because in this
case the above approximations are least critical. In a more detailed
paper~\cite{longpaper} we will present a self-consistency argument to justify
our approach. Figs.~\ref{fig2}-\ref{fig4}  compare our analytical and the exact
(numerically determined) wave functions for low-order excited states and
illustrate the increasing deviation from the fixed wavelength approximation for
more highly excited states. As in the Thomas-Fermi solution in Fig.~\ref{fig1},
the omission of the kinetic energy for the outer regions of the condensate
leads to the artificial truncation at the outer edge~$x=X$ of the wave
functions rather than a Gaussian tail. Also Fig.~\ref{fig4} shows that the
fixed wavelength approximation of~\eq{bec_tf3} for the oscillatory segment of
the solution breaks down for high excitation numbers. Interestingly we have
found that the overlap between the numerical and the analytical solutions to
$Y_4$ is higher for larger values of $\lambda$ (e.g. $\lambda=1/20$). We
believe this is a consequence of the fixed wavelength approximation, we will
explore this point elsewhere.
\newpage
These excited stationary states may soon be realized experimentally. Various
possibilities to generate excited states have been discussed in the
literature~\cite{yyb,Dum98,Dobrek99}. Moreover, Burger {\it et al.}
\cite{Burger99} recently produced a slowly-moving dark soliton using the `phase
imprinting method'~\cite{Dobrek99,Burger99}.  It should be possible with a
slight modification of their experiment to generate the first excited state
$Y_1(x)$ as a {\it stationary} dark soliton.  The generation of higher excited
states should also be possible in a similar manner.

In conclusion, we have derived simple, approximate analytic expressions of the
excited stationary solutions of the 1-dimensional Gross-Pitaevskii equation
using an ansatz based on piecewise solutions. Our analytic solutions agree very
well with numerical solutions for a sufficiently small aspect ratio.
Generalizations of the approach presented here should be rather
straightforward, for example, one can introduce a coordinate dependent
wavelength of the oscillatory part of the solution rather than using the fixed
wavelength approximation. Further details will be explored elsewhere
\cite{longpaper}.

We are grateful to D. Richards and A. Chefles for helpful discussions. This
work was supported by the En\-gineer\-ing and Physical Sciences Research
Council (EPSRC).

\end{multicols} 
\end{document}